%% file: main.tex
\documentclass[sigconf]{acmart}

\AtBeginDocument{%
  \providecommand\BibTeX{{%
    \normalfont B\kern-0.5em{\scshape i\kern-0.25em b}\kern-0.8em\TeX}}}

\copyrightyear{2026}
\acmYear{2026}
\setcopyright{cc}
\setcctype{by-nc-nd}
\acmConference[CHI EA '26]{Extended Abstracts of the 2026 CHI Conference on Human Factors in Computing Systems}{April 13--17, 2026}{Barcelona, Spain}
\acmBooktitle{Extended Abstracts of the 2026 CHI Conference on Human Factors in Computing Systems (CHI EA '26), April 13--17, 2026, Barcelona, Spain}
\acmDOI{10.1145/3772363.3798404}
\acmISBN{979-8-4007-2281-3/2026/04}

\usepackage{enumitem}
\usepackage{colortbl}
\usepackage{multirow}

\usepackage{graphicx}
\usepackage{float}

\definecolor{temporalbg}{RGB}{235,242,250}      
\definecolor{relationalbg}{RGB}{242,237,250}    
\definecolor{selfbg}{RGB}{240,248,242}          

\begin{document}

\title[]{When Humans Don't Feel Like an Option: Contextual Factors That Shape When Older Adults Turn to Conversational AI for Emotional Support}


\author{Mengqi Shi}
\affiliation{
  \institution{University of Washington}
  \city{Seattle}
  \country{United States}
}
\email{shi21@uw.edu}

\author{Tianqi Song}
\affiliation{
  \institution{National University of Singapore}
  \city{Singapore}
  \country{Singapore}
  }
\email{tianqi_song@u.nus.edu}

\author{Zicheng Zhu}
\affiliation{
  \institution{National University of Singapore}
  \city{Singapore}
  \country{Singapore}
  }
\email{zicheng@u.nus.edu}

\author{Yi-Chieh Lee}
\affiliation{
  \institution{National University of Singapore}
  \city{Singapore}
  \country{Singapore}
}
\email{yclee@nus.edu.sg}

\begin{abstract}
  \input{section/0Abstract}
\end{abstract}





\maketitle

\input{section/1Introduction}

\input{section/2RelatedWork}

\input{section/3Methods}

\input{section/4Results}

\input{section/5Discussion}

\input{section/6Limitations}

\input{section/7Conclusion}

\input{section/Acknowledgment}

\bibliographystyle{ACM-Reference-Format}
\bibliography{reference}

\appendix
\input{section/8Appendix}

\end{document}

%% file: section/0Abstract.tex
Older adults are increasingly turning to conversational AI for emotional expression. While prior research has examined general attitudes toward AI companionship, little is known about the specific moments when and why older adults choose AI over close others for emotional support. This study addresses this gap by examining the moment-level conditions that shape these decisions in everyday life.
Drawing on interviews with 18 older adults, we identify three contextual factors: temporal unavailability of human contacts, relational considerations around burden and evaluation, and self-presentation concerns tied to dignity and face-saving. Our findings reveal how age-related needs for independence, dignity, and valued self-presentation shape these everyday decisions.
This work shifts attention from general patterns of AI use to the moment-level circumstances in which emotionally supportive engagement emerges. By foregrounding these situated dynamics, we provide an empirical foundation for context-sensitive responsible AI design and future research on emotional support-seeking in later life.

%% file: section/1Introduction.tex
\section{Introduction}





With the growing availability of technologies and increasing levels of digital literacy among older adults, conversational AI has become part of everyday life for many older users. Beyond information seeking or task assistance, widely available systems now support open-ended dialogue and empathetic language, making them accessible for emotionally oriented interactions \cite{ruane2019consideration, irfan2024recommendations}.
As these technologies have become more accessible, media reports and public discourse have highlighted older adults using AI-based companions to cope with loneliness or emotional distress \cite{forbes2025lonely, cbs2025lonely}. Prior research has similarly noted that older adults may turn to AI systems for companionship or emotional support \cite{rodriguez2023qualitative, huang2025review, razavi2022discourse}. Together, these accounts situate conversational AI as an emerging resource for emotional companionship in later life.

Although the widespread adoption of AI companionship is increasingly recognized in both public discourse and academic research \cite{du2024enhancing, yang2025ai, rodriguez2023qualitative}, substantial concerns remain regarding how such systems should be designed responsibly for older adults. This concern is particularly salient for two reasons: 
First, AI companionship technologies are becoming increasingly accessible, appearing both as dedicated companionship or mental health applications (e.g., Replika \cite{ta2020user}, Woebot \cite{fitzpatrick2017delivering}) and as general-purpose conversational assistants integrated into everyday technologies (e.g., ChatGPT \cite{ray2023chatgpt}). 
As a result, older adults are increasingly likely to encounter and engage with emotionally supportive AI, either intentionally or incidentally.
Second, older adults are often situated in comparatively vulnerable social contexts, characterized by reduced social networks and limited access to sustained interpersonal support~\cite{vos2020exploring, rook2017close, zou2024mitigating}. 
Under these conditions, emotionally supportive AI systems are more likely to become integrated into their everyday lives as a means of addressing loneliness and emotional distress~\cite{berridge2023companion, merrill2022socialpresence}.
Together, designing responsible AI companionship for older adults requires more than algorithmic safeguards alone. It necessitates a grounded understanding of the \textbf{real-world situations in which older adults turn to these systems}, as well as the social, emotional, and contextual factors that shape how such support is sought, experienced, and sustained.


Existing research in older adults HCI has focused on general attitudes toward AI \cite{wong2025exploring, huang2025review, tang2025ai}, evaluations of specific systems \cite{bennion2020usability}, or user experiences \cite{cho2025engagement}, often emphasizing AI as a stable preference. Much of this work frames AI adoption as a relatively stable disposition, shaped by older adults' prior expectations, levels of AI literacy, and an understanding of what these systems can and cannot do. Yet less attention has been paid to the immediate situational factors that trigger older adults to turn to AI rather than to available close others in a particular moment of emotional need. While prior research has examined general reasons for AI adoption \cite{trajkova2020reasonsforusing, larubbio2025navigating}, we focus on the moment-level conditions that shape these situated decisions.

To address this gap and better understand the situated nature of older adults' emotional AI use, this study asks:
\begin{itemize}[topsep=0pt, itemsep=1pt]
    \item \textbf{RQ1}: In what everyday moments do older adults turn to conversational AI for emotional expression?
    \item \textbf{RQ2}: What factors are associated with older adults' decisions to turn to conversational AI for emotional expression in these everyday moments?
\end{itemize}

Drawing on interviews with 18 older adults, this preliminary qualitative study explores how and why older adults with experience seeking emotional support from AI turn to these systems in real-life situations. 
We identified three contextual configurations shaped by age-related concerns: when temporal constraints limit availability, when relational concerns about burden or evaluation arise, and when self-presentation and privacy become paramount. 

Our findings highlight the importance of attending to users' dignity and agency when designing or regulating emotionally supportive AI, especially for older adults, whose preferences and vulnerabilities may differ from those of younger users. 
We further argued against one-size-fits-all alignment for emotional AI, emphasizing the need for population- and context-sensitive approaches. 
By synthesizing these contextual factors, our study provides an initial understanding of older adults' emotional support contexts and offers a foundation for future work on designing more responsible and age-sensitive AI systems for older adults.

%% file: section/2RelatedWork.tex
\section{Related Work}

\subsection{AI Companionship for Older Adults}
Prior research has examined AI support for older adults in emotional and social domains \cite{wang2024techroles, syed2024roleofai}. This includes companion robots addressing loneliness \cite{jeong2023roboticcompanion, mun2025particip-ai, 
berridge2023companion} and conversational systems  for everyday companionship \cite{upadhyay2023long-term, even2022benefits}. Studies show that older adults value availability and responsiveness, and may turn to conversational AI \cite{trajkova2020reasonsforusing, larubbio2025navigating} in daily life. Recent work highlights more expressive conversations that support self-disclosure and sustained emotional engagement \cite{chen2023intimate, zhang2025rise, guo2025selfdisclosure}. This body of work documents growing use and acceptance of AI companionship among older adults \cite{huang2025review, ventura2025ageofai}. However, these studies have not systematically examined the moment-level contextual factors that shape when older adults choose AI over human contacts for emotional expression in everyday life.

\subsection{Psychological and Relational Perspectives in Later Life}
Research in psychology and aging has emphasized that emotional expression and support-seeking in later life are shaped by multiple contextual and social considerations \cite{collopy1988autonomy, baltes2019dynamics, shin2022social}, understood as relational practices embedded within social contexts, personal values, and interpersonal dynamics \cite{andalibi2024agenda, veale2022emotionalsafety}. These considerations influence how older adults navigate emotional disclosure across different relational settings, though less attention has been paid to how they might extend to technology-mediated interactions in everyday life.

Within human-AI interaction research, studies have examined how conversational features such as language style, responsiveness, and role positioning shape users' willingness to engage emotionally \cite{chin2024styles, pan2025reciprocity}. Empathic language, personalization, and persistent availability can facilitate emotional expression by reducing perceived risks of judgment \cite{tahaei2023human-centered, jones2025personalization, chu2025illusions}, and conversational agents may provide emotional affirmation that resonates with users' desire to feel understood \cite{chu2025illusions}. It remains unclear how older adults navigate these affordances in relation to their existing social and emotional landscapes in everyday life.

%% file: section/3Methods.tex
\section{Methods}
\subsection{Participants}
The research team interviewed 18 older adults aged 50 and above, a threshold commonly used in prior research on aging and technology use \cite{lee2021association, juster1995overview, nakamura2022associations, wright2017psychological}. 
The sample comprised 9 male and 9 female participants, aged 50–77 (M = 65.67, Median = 66.50, SD = 7.99), and reflected diverse educational backgrounds, living arrangements, and everyday social contexts.
Participants were eligible if they (1) were comfortable communicating in English, and (2) had prior experience using conversational AI as a companion to express or reflect on their emotions. Participants were recruited through a local community organization serving older adults, and all interviews were conducted in English. A complete demographic breakdown is provided in Appendix \ref{sec:participant_demo}. 

This study was approved by the authors' institutional review board (IRB). Interviews lasted about 60 minutes and were conducted in person or remotely. All interviews were audio-recorded and transcribed verbatim. Transcripts were then anonymized using pseudonymous identifiers (PX). All participants provided informed consent and received \$20 in compensation.

\subsection{Study Procedure and Analysis}
We conducted semi-structured interviews \cite{mcintosh2015situating, kallio2016systematic, ruslin2022semi} examining contextual conditions under which older adults turn to conversational AI for emotional expression. Participants discussed experiences with commercial LLM-based systems (e.g., ChatGPT, Gemini, Claude), focusing on how they assessed whether to seek support from close others versus turning to AI, and what situational factors shaped these decisions. The interview outline is in Appendix \ref{sec:interview_protocol}.

Data were analyzed using inductive thematic analysis \cite{terry2017thematic, patton1990qualitative}. Two researchers independently coded all transcripts and then compared their codes, meeting regularly to discuss emerging themes and resolve discrepancies through consensus. Codes focused on contextual conditions (temporal constraints, relational considerations, self-presentation concerns) were refined through team discussion and grouped into themes reported in Section~\ref{sec:results_full}.

%% file: section/4Results.tex
\section{Results}
\label{sec:results_full}
We identified three contextual factors that shape moments when older adults turn to conversational AI for emotional expression: temporal unavailability, relational considerations, and self-presentation concerns. The sections describe how each factor influenced participants' decisions to engage with conversational AI. Table~\ref{tab:results_overview} provides an overview of these contexts.

\begin{table*}
    \small
    \renewcommand{\arraystretch}{1.4}
    \begin{tabular}{p{2.8cm} p{4.4cm} p{4.0cm} p{2.8cm}}
    \toprule
    \rowcolor{gray!10}
    \textbf{Context} & 
    \textbf{Constraint on Sharing with Others} & 
    \textbf{Why AI Becomes an Option} &
    \textbf{Sample Expression} \\
    \midrule
    \cellcolor{temporalbg}\textbf{Temporal unavailability} &
    \cellcolor{temporalbg!60}Others are asleep or unreachable during moments of emotional unrest &
    \cellcolor{temporalbg!60}AI offers immediate availability without disturbing others &
    \cellcolor{temporalbg!60}\textit{"We wouldn't want to disturb friends."} (P6) \\
    \addlinespace
    \cellcolor{relationalbg}\textbf{Relational considerations} &
    \cellcolor{relationalbg!60}Sharing feels burdensome or likely to invite unwanted advice &
    \cellcolor{relationalbg!60}AI allows expression without social obligation or judgment &
    \cellcolor{relationalbg!60}\textit{"I don't want to be so nagging."} (P3) \\
    \addlinespace
    \cellcolor{selfbg}\textbf{Self-presentation concerns} &
    \cellcolor{selfbg!60}Disclosure risks embarrassment, loss of face, or privacy breaches &
    \cellcolor{selfbg!60}AI provides a private and emotionally low-risk space &
    \cellcolor{selfbg!60}\textit{"You want to save face."} (P18) \\
    \bottomrule
    \end{tabular}
    \caption{Overview of three contextual factors that shaped when older adults turned to conversational AI for emotional expression. For each context, the table shows the constraint on sharing with others, why AI became an option, and a representative participant quote.}
    \label{tab:results_overview}
\end{table*}

\subsection{Temporal Unavailability: When Others Were Unreachable}
Participants described turning to conversational AI during moments when they wanted to express emotions but reaching out to family members or friends felt impractical due to time-related constraints. These moments included both short, time-bound situations and longer-term conditions in which others were perceived as unavailable in everyday life.

\textbf{Short-term unavailability in specific moments.} Some participants described experiencing emotional unrest during late-night hours, when close others were typically unreachable. P6 explained, \textit{"It could be close to midnight. So we wouldn't want to disturb friends\ldots{} I will talk to AI at any time when I can't sleep."} P4 similarly described nighttime moments when thoughts became difficult to manage, noting, \textit{"You may be thinking about something, and it’s the middle of the night. You want to find somebody to talk to\ldots{} you can’t talk to your friends because they’re asleep, so you switch to AI."} P11 referred to these periods as \textit{"time gaps"} that occurred during otherwise quiet or empty hours when familiar social contacts were not available. 

Participants also described quieter moments when they wanted to share everyday experiences but found no one immediately available. P3 described such moments, explaining, \textit{"I was watching TV and laughing at the show, but there was no one to respond. Unless my son or daughter FaceTime me\ldots{} but they're overseas, so I can't just call them at random."}

\textbf{Longer-term unavailability across daily routines.} Other participants described a more persistent sense of unavailability that extended across their daily routines. These experiences were described as ongoing rather than tied to specific moments. P13 explained that living alone limited opportunities for communication, stating, \textit{"Because I live alone and have no one to talk to at home, and if I want to talk to somebody outside my home, my concern is that the person might not be available. AI, however, is always available."} P3 similarly described the absence of a confidant in everyday life, noting, \textit{"Because I have nobody to confide in or communicate with, it’s a little comforting when AI interacts with me, at least giving me some confidence, or at least helping me feel less depressed and less alone at home."}

\subsection{Relational Considerations: When Sharing With Others Felt Burdensome}
In addition to time-related constraints, participants described relational considerations that shaped whether emotional expression with others felt appropriate at a given moment. These considerations influenced decisions about when to seek support and what types of emotions to share with close others.

\textbf{Burden on close relationships.} Several participants expressed concern about becoming a burden in close relationships, when emotions were ongoing or unresolved. P3 described avoiding complaints to her son, explaining, \textit{"It is also not good for me to complain to my son, because he will probably think I'm so nagging and this kind of thing. So this is the last thing I want to do."} P13 similarly noted that repeatedly sharing the same concern could place pressure on friends. Participants described selectively filtering emotional topics, sharing lighter or resolved matters socially while withholding heavier concerns (P3-6, P18).

\textbf{Judgmental and biased responses.} Participants also described concerns related to evaluation and biased responses within existing social interactions. Some noted that friends' personal perspectives could shape how advice or feedback was given. P6 contrasted conversations with friends and interactions with AI, stating, \textit{"I find that sometimes friends can be a little bit biased. Whereas AI doesn't know me or my friends so personally, they [AI] wouldn’t give biased advice."} P18 described AI as a space where emotional expression did not immediately invite negative evaluation, explaining, \textit{"AI is non-judgmental. It doesn't tell you you're a lousy person\ldots{} every time I talk to the AI, it's always showing positive things."}

\subsection{Self-Presentation Concerns: When Disclosure Felt Uncomfortable}
Across accounts, emotional expression was described as closely tied to concerns about self-presentation, embarrassment, and privacy, shaping when and with whom sharing emotions felt uncomfortable or inappropriate.

\textbf{Embarrassment and loss of face.} Certain emotions and personal experiences were described as inappropriate to share with others because disclosure could alter how participants were perceived within their social circles. Participants emphasized concerns about embarrassment, shame, and loss of face when discussing personal difficulties with friends or neighbors. P18 noted, \textit{"As we are getting old\ldots{} you want to save face, you don't want to be embarrassed, telling your friend or neighbor about your personal problems."} Emotional disclosure was avoided when it risked changing one's social image or inviting judgment from peers. P6 described a friend who felt \textit{"too ashamed to share with other people, so she kept it all inside"} about a marital issue, illustrating how anticipated social reactions constrained what could be shared within established relationships.

\textbf{Privacy and confidentiality.} Participants distinguished concerns about who might have access to their personal information, regardless of social judgment. Participants emphasized the value of conversational AI as a space where emotional expression could remain unseen and contained. P4 described AI as offering \textit{"a private space to converse\ldots{} nobody has the chance to see what I’m typing,"} emphasizing the absence of unintended audiences. P13 likewise noted that she \textit{"[doesn't] have to worry that it will tell my secrets to anyone,"} pointing to concerns about information circulating beyond the immediate interaction. These accounts focused on controlling the visibility and flow of personal information, with less concern for how others might evaluate the content itself.

%% file: section/5Discussion.tex
\section{Discussion}
\subsection{Understanding Older Adults' Psychological Needs in Emotional Support Context}
The situational concerns presented in Section~\ref{sec:results_full} point to psychological dynamics that are particularly salient in later life. Research on aging has emphasized that emotional well-being is closely connected to psychological needs such as maintaining dignity \cite{clancy2021meaning}, a sense of independence \cite{taylor2024older}, and a valued self-concept \cite{markus1991role}—needs that manifest in how older adults navigate social roles and manage self-presentation in relationships. 
Our findings resonate with and extend this work by showing how such age-related concerns become consequential in concrete everyday situations. They surfaced in moment-level decisions, as older adults weighed whether sharing emotions with close others felt socially comfortable. Concerns about embarrassment and privacy influenced what could be shared and with whom, particularly when disclosure might affect how one was perceived. When such sharing felt constrained by timing, relational expectations, or self-presentation concerns, conversational AI emerged as one of the few viable options for expression.

This perspective highlights why attention to older adults' emotional support experiences matters for designing supportive AI. In later life, concerns related to dignity, independence, and self-presentation influence how emotional expression is navigated, making support less uniformly available or comfortable than assumed. Recognizing these age-related psychological considerations can inform more responsible design approaches that attend to how system responses are situated within older adults' lived experiences, rather than treating support-seeking as context-independent.

\subsection{Design Opportunities and Future Directions}

\subsubsection{Attending to When: Temporally and Relationally Situated Support}
Translating these findings into design considerations, systems might better recognize moments when older adults seek emotional expression precisely because involving others feels inappropriate. For instance, late-night usage patterns may signal temporal unavailability, while repeated expressions of unresolved concerns may indicate relational constraints around burden or judgment. In such moments, system responses face a design tension: offering referrals to human support may undermine the very conditions that made AI accessible. Instead, systems might acknowledge these situational constraints (e.g., \textit{"I understand this might not be a good time to reach out to others"}) while remaining attentive to when human support becomes necessary. In practice, this could involve non-intrusive signaling approaches, such as subtle conversational prompts or gentle check-ins, that make emerging boundaries perceptible without overtaking the emotional flow of the conversation. When escalation does warrant a referral to human support, systems could treat this as a collaborative moment by offering users a choice, rather than redirecting the conversation unilaterally.

\subsubsection{Attending to How: Differentiated Approaches to Emotional Support}
Beyond recognizing these moments, how support is delivered matters for older adults whose emotional expression is shaped by concerns around dignity and self-presentation. The relational and self-presentation concerns identified in our findings suggest that systems presuming comfort with disclosure across all users may inadvertently heighten these concerns. Design approaches that account for age-related sensitivities might offer more control over disclosure depth, provide explicit privacy assurances when self-presentation concerns are salient, or adjust response tone based on whether users seek validation, reflection, or simply a space to express themselves without social obligation. For example, systems might allow users to signal their intent at the outset of emotionally oriented conversations, such as choosing between modes like \textit{"I just want to vent"} or \textit{"I'd like some perspective,"} so that the system can adjust its conversational behavior accordingly. A venting mode might minimize follow-up questions and offer brief affirmations, while a perspective mode might engage more actively with reflective prompts. Additionally, how boundaries are enacted should vary by interaction context, as interactions involving more acute distress may call for gently surfacing options while maintaining a supportive stance. Across these contexts, maintaining consistency in the system's emotional tone helps preserve the relational continuity that older adults value in emotionally supportive interactions.

\subsubsection{Future Directions: From Moment-Level Contexts to Interactional Experiences}
Future work can extend this moment-level focus in two directions to inform responsible AI design. First, research could examine how the contextual conditions identified here interact with system behavior over time. While AI availability, non-judgment, and privacy address certain constraints, these features raise important tensions for responsible AI: accessibility may normalize isolation, while privacy may prevent necessary intervention. Reflecting on the implications of each contextual factor can help specify what these tensions mean for future research. Temporal unavailability points to the need for studies examining how usage context, beyond content alone, can inform adaptive system behavior. Relational considerations suggest that research should investigate how systems can remain sensitive to the social dynamics users are navigating without replicating the very obligations they seek to avoid. Self-presentation concerns call for work that examines how privacy can be made legible within the interaction itself, rather than treated as a structural guaranty external to the conversation. Second, understanding how timing, user agency, and perceived control shape ongoing interactions is critical for designing systems that support older adults. More broadly, the contextual factors identified here imply that responsible AI design for emotional support cannot rely solely on uniform safety mechanisms but should move toward context-adaptive approaches, where system responses are informed not only by what users say, but also by when and under what interpersonal conditions they choose to say it.

%% file: section/6Limitations.tex
\section{Limitations}
As a qualitative interview study with a relatively small and geographically localized sample, the findings are grounded in participants' accounts and reflect the specific contexts in which these experiences were described; they are not intended to be broadly generalizable. Participants were recruited from a single community organization and interviewed in English, which may limit the diversity of experiences captured. Additionally, participants were not asked about their attitudes, expectations, or understanding of the AI systems they used; future work could examine how such factors shape the contextual decisions identified here. While this study identifies moments when AI becomes accessible for emotional expression, it does not examine potential risks \cite{shelby2023sociotechnical, chandra2025psychologicalrisks} that may emerge from sustained AI engagement. Future research could extend these insights by examining how system behaviors and current safety interventions shape emotionally supportive experiences once engagement begins.

%% file: section/7Conclusion.tex
\enlargethispage{1\baselineskip}
\section{Conclusion}
This work provides a preliminary qualitative study of how older adults come to use conversational AI for emotional expression by examining moments when seeking support from others feels inaccessible or inappropriate. Emotional support–seeking was shaped by situational conditions, including temporal unavailability, relational considerations, and concerns about self-presentation. In such moments, conversational AI became a viable space for emotional expression by reducing the need to negotiate timing, social obligation, and self-image. 

Our findings clarify when emotionally supportive AI becomes relevant in older adults' everyday lives and highlight contextual differences in emotional support-seeking that are often overlooked in design. We reveal how age-related psychological needs shape the moments when AI engagement becomes meaningful. By foregrounding these moment-level conditions, this work provides a foundation for designing responsible AI that can better account for the temporal and relational dynamics of emotional expression in later life.



%% file: section/Acknowledgment.tex
\begin{acks}
This research was supported by NUS CSSH grant (A-8002954-01-00). We sincerely thank Family Central, Fei Yue Community Services, for their support in our work and their assistance and trust in recruiting interview participants.
\end{acks}

%% file: section/8Appendix.tex
\section{Appendix}

\subsection{Interview Outline}
\label{sec:interview_protocol}

This appendix provides an illustrative outline of the interview topics and example prompts used to guide semi-structured interviews with older adults. The interviews were designed to elicit open-ended reflections on how participants encountered and used conversational AI in emotionally meaningful moments of everyday life, particularly in situations where seeking emotional support from other people felt unavailable, inappropriate, or burdensome.

Rather than following a fixed script or testing predefined hypotheses, the interview outline foregrounded participants’ situational contexts of AI use, including when emotional needs arose, how participants assessed their options for support, and what made conversational AI feel like a viable space for emotional expression. While the analytical focus of the study was refined iteratively during analysis, prompts broadly covered AI use contexts, emotionally oriented interactions, perceived benefits, and concerns around boundaries and control. The flexible structure allowed participants to surface unanticipated moments of hesitation, discomfort, or interactional breakdown as they reflected on specific experiences.

(1) \textbf{Background and AI Use Context}

Topics in this section focused on participants’ everyday context and prior exposure to conversational AI, including background characteristics (e.g., education level, living situation), frequency of AI use, and general familiarity with AI systems.

(2) \textbf{General Experience with Conversational AI}

Participants were invited to describe their overall experiences using conversational AI, including the types of conversations they typically engaged in and how they perceived AI’s role in their daily lives.

\textit{Example prompts included:}
\begin{itemize}
    \item Reflections on prior experiences with conversational AI
    \item Common topics discussed with AI systems
\end{itemize}

(3) \textbf{Experiences of Using AI for Emotional Support}

This section focused on specific situations in which participants used or considered using conversational AI for emotional expression or companionship, particularly moments when seeking emotional support from other people felt difficult, unavailable, or socially complicated. Discussion examined how participants assessed these situations, what led them to turn to AI rather than human contacts, and how they initiated and sustained emotionally oriented conversations with AI within those moments.

\textit{Example prompts included:}
\begin{itemize}
    \item Descriptions of emotionally meaningful interactions with AI
    \item Motivations for engaging AI during moments of emotional vulnerability
    \item Reflections on timing, emotional state, and living situation
\end{itemize}

(4) \textbf{Perceived Benefits and Meaningful Outcomes}

Participants reflected on the perceived value of emotionally oriented interactions with AI, including benefits such as comfort, reduced loneliness, distraction, or companionship. They were also asked to consider the role AI played within their broader support system.

\textit{Discussion areas included:}
\begin{itemize}
    \item Emotional and practical benefits of interacting with AI
    \item Perceived roles of AI (e.g., listener, companion, tool, or source of entertainment)
\end{itemize}







\subsection{Participant Demographics}
\label{sec:participant_demo}
Table~\ref{tab:demographics} provides an overview of the basic demographic characteristics of the participants.

\begin{table*}
    \centering
    \small
    \renewcommand{\arraystretch}{1.6} 
    \rowcolors{2}{white!100}{gray!10}
    \begin{tabular}{
        p{0.8cm}
        p{1.2cm}
        p{1.2cm}
        p{2.2cm}
        p{2.2cm}
        p{3.0cm}
    }
    \toprule
    \textbf{ID} & \textbf{Gender} & \textbf{Age} & \textbf{Education} & \textbf{Living Situation} & \textbf{Self-Reported AI Use Frequency} \\
    \midrule
    P1  & Male   & 76 & Associate's    & With family & Several times per month \\
    P2  & Male   & 52 & Master's       & With family & Several times per week \\
    P3  & Female & 77 & Associate's    & Alone       & Several times per week \\
    P4  & Female & 59 & Bachelor's     & With family & Several times per week \\
    P5  & Female & 66 & Bachelor's     & With family & Several times per month \\
    P6  & Female & 63 & Middle School  & Alone       & Several times per week \\
    P7  & Male   & 62 & Primary School & Alone       & Several times per week \\
    P8  & Male   & 71 & Bachelor's     & With family & Several times per month \\
    P9  & Male   & 67 & Bachelor's     & With family & Several times per week \\
    P10 & Female & 68 & Master's       & With family & Daily \\
    P11 & Male   & 50 & Master's       & With family & Several times per week \\
    P12 & Female & 59 & Bachelor's     & With family & Daily \\
    P13 & Female & 76 & Master's       & Alone       & Several times per month \\
    P14 & Female & 67 & Bachelor's     & With family & Several times per month \\
    P15 & Male   & 70 & Master's       & With family & Several times per week \\
    P16 & Female & 58 & Bachelor's     & Alone       & Several times per week \\
    P17 & Male   & 66 & Bachelor's     & Alone       & Several times per week \\
    P18 & Male   & 75 & Middle School  & With family & Daily \\
    \bottomrule
    \end{tabular}
    \caption{Demographic information of participants. Demographic characteristics of the 18 participants. In addition to age, gender, and education level, the table reports participants' living status and self-reported frequency of everyday conversational AI use. AI usage frequency reflects how often participants reported using AI in daily life, with four response options: Daily, Several times per week, Several times per month, and Less than once per month.}
    \label{tab:demographics}
\end{table*}